\documentstyle[prl,aps,epsfig]{revtex}
\draft
\begin{document}
\twocolumn[
\hsize\textwidth\columnwidth\hsize\csname@twocolumnfalse\endcsname
\hskip 2cm\vskip -1cm ETH-TH/99-13
\title{Critical Ising modes in low-dimensional Kondo insulators}
\author{Karyn Le Hur}
\address{Theoretische Physik, ETH-H\"onggerberg, CH-8093 Z\"urich, Switzerland}
\maketitle

\begin{abstract}
We present an Ising-like intermediate phase for one-dimensional Kondo
insulator systems. Resulting from a spinon splitting, its low-energy 
excitations are critical Ising modes, whereas the triplet sector has a 
spectral 
gap. It should occur as long as the RKKY oscillation amplitude dominates over
any direct exchange between localized spins. The chiral fixed point, however,
becomes unstable in the far Infra-Red limit due to prevalent 
fluctuations among 
localized spins which induce gapless
triplet excitations in the spectrum. Based on previous numerical results, we
obtain a paramagnetic disordered state ruled by the correlation length of the
single impurity Kondo model. 
\end{abstract}
\twocolumn
\vskip.5pc ]
\narrowtext

\section{Introduction}

The one-dimensional (1D) half-filled Kondo lattice is a simple model for a
group of compounds called ``the Kondo insulators''\cite{Aeppli}. They exhibit
the high-temperature behavior of usual Kondo systems, such as the
Curie-Weiss-like magnetic susceptibility, but
at lower temperatures evolve into a semiconducting
phase with small gaps. At low-energy, a
Kondo insulator is a typical realization of spin-charge
separation. This aspect manifests itself in the difference in size between the
charge gap and spin gap. 

Since its discovery, the indirect
Ruderman-Kittel-Kasuya-Yosida (RKKY) interaction between localized magnetic
impurities embedded in a host metal has played an important role in the theory
of magnetism. In 1D the RKKY oscillation amplitude displays only a
very slow decay $\sim 1/x$. The conduction electrons are subject to the 
resulting exchange field which oscillates spatially
with the Fermi wavelength. It produces Kondo localization. This conclusion
was obtained by an exact diagonalization study\cite{Tsunetsugu}. The result
has been confirmed with the density matrix renormalization group 
method\cite{Yu}, and later supported by the
mapping to a non-linear sigma model\cite{Tsvelik} and by the bosonization
approach\cite{Zachar,KLH}. The charge gap varies linearly with the Kondo
coupling\cite{Shibata}. In this paper another important question is examined:
given the RKKY oscillation, what are the theoretical consequences for
the low-temperature spin properties?

The RKKY interaction (as we shall see later) cannot break the spin
rotational
symmetry due to Kondo localization. Fluctuations in 
the local spin configurations lead to
a paramagnetic fluid. Since the model
for the local spin degrees of freedom is not purely integrable, no
exact solution has been proposed to date. In fact, by considering an
RKKY-like spin interaction, a topological term 
{\it \`a la Haldane}\cite{Haldane0} is not
justified so that gapless magnons with S=1 should be the
prevalent excitations over the spin array. In consequence, 
the spin effective model we shall study may be viewed 
at intermediate distances as an
S=1/2 Heisenberg chain interacting with a `quasi-ordered' spin
array (with non-topological configurations). Conventionally, the low-lying 
excitations (often called
spinons) of the S=1/2 Heisenberg spin chain are represented by one gapless
bosonic mode. In the language of the theory of critical phenomena, this means
that the model is described by a central charge which is equal to C=1.
For the present Kondo system, we will show how
the frustration imposed by the localized spins may play its tricks and the
bosonic mode or spinon decouples into two modes of Majorana fermions (with
C=1/2 each) having different spectra. The spinon splitting phenomenon is at
the basis of the new chiral fixed point presented here and also
occurs in spin ladders which retain the local
symmetry of the Kagom\'e net\cite{Azaria}. Note, however, that 
in our problem including the 
dynamics of localized spins will make the chiral fixed point unstable in the 
far Infra Red (IR) limit. In
agreement with previous numerical results\cite{Tsunetsugu,Shibata}, the
system should flow to a disordered spin liquid phase with a finite
antiferromagnetic (AFM) length scale $\xi_{AFM}$ close to that in the
single impurity Kondo model. In the following, we shall introduce the Kondo
spin liquid phase as a direct consequence of the chiral fixed point
instability in the presence of strong fluctuations. 
\vskip 0.1cm
The starting point is the Hamiltonian:
\begin{equation}
{\cal H}=-t\sum_{\langle i,j\rangle, \alpha} 
c^{\dag}_{i,\alpha}c_{j,\alpha}+h.c. + J_K\sum_{i,\alpha,\beta} 
c^{\dag}_{i,\alpha}\vec{\sigma}_{\alpha,\beta}c_{i,\beta}.\vec{S}_i
\end{equation}
The first term represents the electron hopping between nearest-neighbor sites
$i$ and $j$. The second term is the Kondo coupling $(J_K\ll t)$ between the 
localized
spin $\vec{S}_i$ and the mobile electron at the same site. Adding a direct
exchange $J_H>0$ between the nearest {\it core} spins gives quite
different physics. It ensures the presence of local AFM fluctuations
leading to topological configurations for the localized spins. If the relation
$J_H\gg J_{RKKY}\sim {J_K}^2/t$ is satisfied, the spin theory at low 
temperature
is described in terms of an O(3) nonlinear $\sigma$ model where the
topological term has no contribution\cite{Tsvelik,Betouras}. That produces
disordered Kondo fluids with quite short AFM correlation lengths
$\xi_{AFM}\simeq \exp(\pi S)$. Excitations of the O(3) nonlinear $\sigma$
model are S=1 triplets. In the extreme limit $J_H\gg J_K$, the physics
becomes similar to the two-leg spin ladder\cite{Rice}; spinons confine to 
form  both S=1 triplet and singlet excitations with gaps $m_t$ and $m_s$ 
$(m_t,m_s\propto J_K^2$ and $\xi_{AFM}\sim m_t^{-1}$)\cite{White,KLH2}. 

Here, we mainly focus on the interesting case $J_H\rightarrow 0$.

\section{Basis of our formalism}

The approach followed in this paper is based on bosonization techniques
for both charge 
and spin degrees of freedom of the conduction electrons. For complete
reviews, see refs.\cite{Haldane,affleck}. 

For $J_K=0$, the model for the
conduction band is gapless and is characterized by the separation of spin
and charge. Its low-energy spin properties belong to the same universality 
class as the Heisenberg model. The low-temperature behavior is then
described by the level-1 SU(2) Wess-Zumino-Witten (WZW) conformal field theory
(CFT)\cite{Tsvelik2}.  
The physical particles (or spinons=spin 1/2 excitations)
are included through the primary 
fields $\Phi^{(1/2)}$ and $\Phi^{(1/2)\dag}$ from the representation of the 
$SU(2)$ group\cite{Bernard}. The WZW action taking into account the dynamics of 
the spinon
objects is explicitly given by:
\begin{eqnarray}
& &S_{WZW}=-\frac{1}{16\pi}\int d^2 x
Tr(\partial_{\mu}\Phi^{(1/2)\dag}\partial_{\mu}\Phi^{(1/2)})\\ \nonumber
&+&\frac{i}{24\pi}\int_0^{\infty} d\epsilon\int d^2 x
\epsilon^{\alpha\beta\gamma} Tr({\cal A}_{\alpha}{\cal A}_{\beta}{\cal A}_
{\gamma})\\
\end{eqnarray}
and, ${\cal A}_{\alpha}=\Phi^{(1/2)\dag}\partial_{\alpha}\Phi^{(1/2)}$.
This description, which has its origin in the structure of the Haldane-Shastry
spin chain with $1/x^2$ exchange\cite{HS}, stresses the fact that the 
fundamental
fields in this theory (or spinon fields) may be viewed as free fields apart
from purely statistical (in this case: semionic\cite{Bouwknegt}) interactions 
that may be
taken into account by a rule generalizing the Pauli principle.
The electronic spin density is represented as
$\vec{S}_c(x)=\vec{J}_c(x)+e^{i2k_Fx}\vec{n}_c(x)$, where
$\vec{J}_c=\vec{J}_{cR}+\vec{J}_{cL}$ and 
\begin{equation}
\vec{n}_c=\frac{1}{2\pi
a}\hbox{Tr}\{\vec{\sigma}(\Phi^{(1/2)}+\Phi^{(1/2)\dag})\}\cos{\sqrt{2\pi}
\Phi_c} 
\end{equation}
are,
respectively, the smooth and staggered parts of the magnetization. 
The precise relationship between the chiral spin currents and the
spinon fields has been discussed in detail in ref.\cite{affleck}.
The lattice
step $a$ defines the required short-distance cut-off.

The charge sector
is similarly described in terms of a U(1) scalar field $\Phi_c$ leading to a 
CFT
central charge C=1 for holons\cite{BA} as well. In 
the charge sector, the free action
is given by:
\begin{equation}
S_{g;v_{\rho}}=\frac{1}{2}\int dxd\tau\
\frac{1}{v_{\rho}g}(\partial_{\tau}\Phi_c)^2-\frac{v_{\rho}}{g}(\partial_x
\Phi_c)^2
\end{equation}
Neutral excitations are 1D acoustic 
plasmons which propagate
with velocity $v_{\rho}$ and are characterized by the Luttinger parameter
$g=v_F/v_{\rho}$. For a 1D free electron gas $g$ is equal to one and  
$v_F=2t\sin(k_Fa)$ is the velocity for the charge and spin degrees of 
freedom. In the continuum limit the Kondo interaction takes the form:
\begin{eqnarray}
\label{zero}
{\cal H}_{int}&=&\lambda_2(\vec{J}_{cL}+\vec{J}_{cR})\vec{S}_j\\ \nonumber
&+&\frac{\lambda_3}{2\pi a}e^{i(2k_F-\pi)x}\ \hbox{Tr}(\vec{\sigma}
\Phi^{(1/2)}).\cos\sqrt{2\pi}{\Phi}_c\vec{S}_j+h.c.
\end{eqnarray}
where $\lambda_{2,3}\propto J_K$. Note that $\delta=2k_F-\pi$ measures the 
deviation
from half-filling. Here, we mainly consider the half-filled case
$\delta\rightarrow 0$. The configuration of localized spins can be also
parameterized as 
\begin{equation}
\vec{S}_j=S(a\vec{L}_j+(-1)^x
[1-(a{L}_j)^2]^{1/2}\vec{n}_j)
\end{equation}
where $j$ enumerates the sites and
$\vec{L}_j.\vec{n}_j=0$. Finally, to quantify correctly the local spins, we
must use the path integral representation. It leads to an extra term in the
action (the so-called Berry phase)\cite{Fradkin,Tsvelik}:
\begin{equation}
S_{B}=iS\int dxd\tau\ \vec{L}.(\vec{n}\times \partial_{\tau}\vec{n})
\end{equation}
As already mentioned in the introduction, the RKKY interaction between local
moments should play a crucial role, for $J_K\ll t$.

\section{RKKY interaction as a staggering field}

The standard treatment of the 
RKKY interaction corresponds to a calculation of the correlation
function 
\begin{equation}
C(x)=\frac{\cos(2k_Fx)}{2\pi a}\langle S_0^z
\hbox{Tr}{\sigma}^z\Phi^{(1/2)}(x)\cos\sqrt{2\pi}{\Phi}_c(x)\rangle
\end{equation}
 The 
angular brackets
indicate a thermal average. This function describes the spatial correlation of
the electron-spin density with the impurity spin on $j=0$. Another impurity 
spin located at $j=x$ would see this correlation, and lowest-order perturbation
theory in $\lambda_3$ constitutes an exact derivation of the RKKY law. For $
x<\xi_T=v_F/\hbox{k}_B T$, this yields: $H_{RKKY}=\lambda_{RKKY}\sum_{\langle
0,x\rangle}\vec{S}_0\vec{S}_x$
with
\begin{equation}
\label{trois}
\lambda_{RKKY}=\frac{\lambda_3}{2\pi a}C(x)=\frac{-\lambda_3^2}{2\pi v_F}
\frac{\cos(2k_Fx)}{x^g}a^{-1}
\end{equation}
In the range of temperature $x<\xi_T$, the RKKY oscillation amplitude
displays only a very slow algebraic decay. In the noninteracting case $g=1$,
the usual $x^{-1}$ decay is recovered.
The RKKY law, prominent at 
high temperatures,
should then favor the following configuration for the localized 
spins: $\vec{n}_j=\frac{1}{a}\vec{n}$ where $\vec{n}$ is a {\it fixed} unit 
vector and 
$1-(a{L}_j)^2\sim 1$. Using the Hamiltonian (\ref{zero}), we find that it 
generates a static staggering magnetic field 
$h_s=S\lambda_3$ on conduction electrons: 
\begin{eqnarray}
{\cal H}_{int}&=&h_s\vec{n}_c\cdot\vec{n}\\ \nonumber
&=& \frac{h_s}{2\pi a^2}\cos\sqrt{2\pi}{\Phi}_c\hbox{Tr}(\vec{\sigma}
\Phi^{(1/2)})\cdot\vec{n}+h.c.
\end{eqnarray}
At high temperatures, the conduction electrons are subject to a perfectly
static potential as if there was a finite staggered spin moment.

Now, we expand the partition function ${\cal Z}={\cal Z}_0+\delta{\cal Z}
[{h_s\neq 0}]$
to the second order in $h_s$. We find that 
$\delta {\cal Z}/{\cal Z}_0$ is equal to\footnote{To simplify the
expression, we have taken $v_{\rho}\sim v_F=1$}:
\begin{equation}
\hbox{\Large(}\frac{hs}{2\pi}\hbox{\Large)}^2
\int \frac{d\tau_1 d\tau_2 dx_1 dx_2}{a^4}\ \hbox{\Large(}
\frac{(x_1-x_2)^2+(\tau_1-\tau_2)^2}{a^2}\hbox{\Large)}^{-\frac{1+g}{2}}
\end{equation}
${\cal Z}_0$ is the partition function of the free system with $J_K=0$. 
To make the (total) partition function 
invariant under the cut-off transformation $a\rightarrow a'=ae^{d\ln L}$, 
$h_s$ has to obey: 
\begin{equation}
\frac{dh_s}{d\ln L}=(2-\frac{1}{2}-\frac{g}{2})h_s
\end{equation}
Note that ${\cal Z}_0$ is not affected by such a rescaling.
Starting with a free electron gas
$(g\rightarrow 1)$, we confirm that it produces
the localization of conduction electrons at an energy scale $\Delta_c\propto 
h_s=J_K/2$\cite{Tsvelik,Shibata}, defined such as $h_s(\Delta_c)\sim 1$. 
It should be noted that
spin flip events do not contribute to the perturbative 
result (\ref{trois}), to this order. Of course, the exponent $g$ is
also affected by the Kondo interaction. We find:
\begin{equation}
\frac{dg}{d\ln L}=-2\pi g^2\lambda_3^2 J_o(\delta(L)a)
\end{equation}
$J_o$ is the Bessel function. Since we restrict our arguments to
the case of half-filling, we put $\delta=0$ and $J_o(0)=1$.
Correlations of $\cos\sqrt{2\pi}{\Phi}_c$ show a long-range
order at zero temperature because the renormalized exponent $g^*$ goes
to zero. We can average
$\langle\cos\sqrt{2\pi}{\Phi}_c(x)\rangle\sim\sqrt{\Delta_c}$. The charge
motion is frozen and the leading order parameter in the 
charge sector is the so-called
$4k_F$ charge density wave (CDW): $\rho_{4k_F}\propto \cos
\sqrt{8\pi}{\Phi}_c(x)$. As for the 
occurrence of the Mott-Hubbard gap due to Umklapps\cite{Haldane}, 
the (Mott)-Kondo
insulating state occurs due to the pinning of the $4k_F$ CDW\cite{remark}:
\begin{equation}
\langle \rho_{4k_F}(x)\rho_{4k_F}(0)\rangle \propto x^{-4g^*}\sim 
\hbox{constant}
\end{equation}
The charge field, on the other hand, could also be affected by
the disorder. As shown in ref.\cite{KLH}, a strongly disordered 1D Kondo array
will crossover to an Anderson localization state. Thus, we now study
spin properties of such an insulating (Mott)-Kondo state with very small
randomness. 

Since $\lambda_3$ violates the separation of charge and spin one
can expect massive triplet modes with a spectral gap 
$J_KS$. At quite short distance, one can treat $\vec{n}$ as a constant
vector. Writing $\Theta=i\vec{\sigma}.\vec{n}$, the result is a uniaxial Kondo
effect, leading to an Ising-like solution $\Phi^{(1/2)}=\Phi^{(cl)}$. 
Using properties of Pauli matrices\cite{Betouras,Shelton}:
\begin{equation}
\hbox{Tr}\vec{\sigma}\Phi^{(cl)\dag}\hbox{Tr}
\vec{\sigma}\Theta+h.c.=\hbox{Tr}\Phi^{(1)},\ 
\Phi^{(1)}=\Phi^{(cl)\dag}\Theta
\end{equation}
 we check that spins of
conduction and localized electrons confine to form triplets (or magnons
represented by the `composite' field $\Phi^{(1)}$) with a mass $m_t=
\Delta_c$ . Such a classical solution is ruled by: 
\begin{equation}
\vec{n}_c=
\hbox{Tr}\vec{\sigma}\Phi^{(cl)\dag}=-\hbox{Tr}\vec{\sigma}\Theta=-\vec{n}
\end{equation}
When $x\simeq {\cal L}_{loc}$ with ${\cal L}_{loc}=(J_K S)^{-1}$, the
so-called spin density glass state aims to arise\cite{KLH}: conduction
electrons feel a staggered potential within this scale and open a
quasiparticle gap due to Bragg scattering.
However, starting with an isotropic system, it cannot be longer stabilized for
temperatures much lower than $\Delta_c$. Indeed, from Eq. (\ref{trois}), we
find the following recursion law for the RKKY exchange in the
delocalized phase: 
\begin{equation}
\frac{d\lambda_{RKKY}}{d\ln L}=\frac{1}{2\pi v_F}{\lambda_3}^2
\end{equation}
It cannot provide an energy scale more relevant 
than $\Delta_c$ since it involves processes in ${\lambda_3}^2$ (
the electron gas localization occurs due to scattering events $\propto
\lambda_3$). Then, the conduction electrons
effectively `screen away' the internal field before a true magnetic transition
(with breaking SU(2) symmetry) can occur. 

\section{SU(2) symmetry not broken: consequences}

The uniaxial solution 
$\Theta(x,\tau)=\Theta$ and
$\Phi^{(1/2)}=\Phi^{(cl)}$ is not available at long distances: {\it there are
certainly states in the gap}. 

For $J_H\gg J_{RKKY}$, it is important to include a topological term (for
the local moments) first derived by Haldane\cite{Haldane0}:
\begin{equation}
\label{to}
S_{top}=\frac{i}{8\pi}\int dxd\tau \epsilon_{\mu\nu}\hbox{\large(}\vec{n}
\cdot[
\partial_{\mu}\vec{n}\times\partial_{\nu}\vec{n}]\hbox{\large)}
\end{equation}
Then, the model becomes solvable in the semi-classical limit 
(`large-S expansion'), by including the Berry phase and integrating out
both conduction electron variables and fast modes $\vec{L}$. The result is
an 0(3) nonlinear $\sigma$ model built out
the field $\vec{n}(x,\tau)$ (with 
triplet excitations). The resulting topological term has no 
contribution and then
the gapless ordered state of the isotropic
sigma model is marginally unstable (due to the implicit breaking
of conformal invariance) and it opens a gap (`the so-called Haldane gap')\cite
{Haldane0,Tsvelik}:
\begin{equation}
m\sim J_K S\exp -\pi S
\end{equation}
which is not so far from $\Delta_c$. 
The spin gap $m$ has a topological origin, and 
excitations in the interval between 
$m$ and $J_K S$ are known to
be massive spin polarons formed due to an interaction 
between electrons and kinks of the unit vector 
$\vec{n}(x,\tau)$\cite{Tsvelik}. To my knowledge, such Kondo insulator 
(ruled by two different energy scales in charge and spin sectors)
was the first realization of so-called spin-charge separation. 
On the other hand, a
direct exchange between localized moments is crucial for the occurrence of
massive spin polarons\cite{Betouras} and such a topological description
is not expected
to remain available when $J_H=0$. 
Numerical results\cite{Tsunetsugu,Shibata} predict in
contrast a pure Kondo ground state with a very large AFM length 
$\xi_{AFM}$ driven by 
spin flip processes or the Kondo term $\lambda_2$, which
has not been taken into account in the semi-classical approach of
ref.\cite{Tsvelik}. 
To obtain precisely low-energy spin excitations in this limit we adopt
the following scheme. 

For $J_H=0$, there is no mathematical justification 
to include a topological term of the form of Eq.(\ref{to}) for
local moments since the model is not integrable. Then we
do not introduce it. In consequence, we start
with a spin array which is 
described by usual spin waves with spin ${\Delta
S}^z=\pm 1$. Since a true magnetic transition does not arise,
ferromagnetic local fluctuations may occur in the
far IR limit leading to {\it free} triplet particles in the spectrum.
The forward Kondo scattering term should then play a crucial role
at very long distances. The effect of
fast variables $L_j$ will be considered later 
when $x\sim \xi_{AFM}$ (i.e. when the presumed quasi long-range order
is completely destroyed). Using the so-called operator
product expansions of ref.\cite{Fujimoto}, we can check that 
$\lambda_2$ is not renormalized by a term like $g^*{\lambda_3}^2/2\pi v_F$
due to the {\it insulating} nature of the ground sate ($g^*\rightarrow
0$). Ferromagnetic fluctuations are well decoupled if we
neglect the Berry phase (which is only reponsible for the
correct quantization of local spins).

Before, let us show that spinons in the electron gas
cannot {\it vanish} (confine) completely when the spin array
develops only a {\it quasi-order} (with isotropic
correlation functions): the spin density glass state becomes
less stable and critical Ising modes can survive in the spectrum. 

\subsection{Refermionization and critical Ising modes}

Let us start with a spin array which satisfies the following quasi-order
condition: 
\begin{equation}
\langle
n^z(x)n^z(0)\rangle=\langle n^{+}(x)n^{-}(0)\rangle\propto x^{-2\alpha}
\end{equation}
It means that the local magnetization operator $n^i(x,\tau)$ with $i=x,y,z$ has
the scaling dimension $\alpha$. Since the
RKKY interaction between local moments yields only
a very slow algebraic decay in 1D, we deduce: 
$\alpha\ll 1/2$, 1/2 being the scaling dimension of the staggered magnetization
operator in the Heisenberg chain with nearest neighbor exchange\cite{affleck}).
To solve the problem at intermediate
distances (when
${\cal L}_{loc}\ll x\ll \xi_{AFM}$), we assume that the parameter
$\alpha$ is too small that we can replace in first
approximation local moment operators 
by effective expectation values
$\langle n^z(x,\tau)\rangle\sim\langle n^{\perp}(x,\tau)\rangle=
\gamma\neq 0$,
obtained by averaging in time and space slow fluctuations in
the spin array. The electron gas is submitted to an SU(2)-invariant 
{\it quasi-static} staggering potential.
Then, it is not difficult to integrate out
local spin degrees of freedom.
To treat correctly the electron gas, it is now convenient to use the 
Abelian representation\cite{KLH2},
\begin{equation}
\vec{n}_c\sim(\cos\sqrt{2\pi}\Theta_s,\ -\sin\sqrt{2\pi}\Theta_s,\ 
\sin\sqrt{2\pi}\Phi_s)
\end{equation}
where $\Theta_s$ is the field dual to $\Phi_s$. Then, 
at long distances, the term
$\lambda_3$ can be symmetrized as\cite{KLH}:
\begin{eqnarray}
\label{deux}
{\cal H}_{int}&\sim&\frac{1}{2\pi a}\Delta_c^{3/2}(\sin\sqrt{2\pi}\Phi_s(x)+
\sin\sqrt{2\pi}\Theta_s(x))\\ \nonumber
&\sim& \frac{\Delta_c}{2\pi a}(\cos\sqrt{4\pi}\Phi_s(x)+\cos\sqrt{4\pi}
\Theta_s(x))
\end{eqnarray}
We have replaced the charge operator by its expectation value and
used the definition of $\Delta_c=J_K S$.
In view to respect conventional notations we have changed $\Phi_s\rightarrow
\Phi_s+\sqrt{\pi}/4$ and $\Theta_s\rightarrow \Theta_s+\sqrt{\pi}/4$ in the
second equation. In ref.\cite{KLH}, we did not solve the Hamiltonian in
the isotropic case. To solve it, we require an effective fermionic theory 
similar to Hubbard\cite{Schulz} or spin ladder\cite{Shelton} models, and
carbon nanotube problems\cite{Egger}. 
The refermionization technique is defined, as follows.

Let us define new effective fermion 
operators for the
spin channel. Right- and left-moving components $(q=\pm=R,L)$ can be written
in terms of the bosonic phase field,
\begin{equation}
\psi_{sq}(x)=\frac{\eta_q}{\sqrt{2\pi a}}\exp
\large\{-i \sqrt{\pi}(q\Theta_s+\Phi_s)(x)\large\}
\end{equation}
Then we have
\begin{eqnarray}
\label{un}
\frac{1}{\pi a}\cos\sqrt{4\pi}\Phi_s&=&\eta_R\eta_L(\psi_{sR}^{\dag}\psi_{sL}-
\psi_{sL}^{\dag}\psi_{sR}),\\ \nonumber
\frac{1}{\pi a}\cos\sqrt{4\pi}\Theta_s&=&-\eta_R\eta_L(\psi_{sR}^{\dag}
\psi_{sL}^{\dag}-\psi_{sL}\psi_{sR}).
\end{eqnarray}
Klein factors have been chosen as $\eta_R\eta_L=i$.
Apart from the usual mass bilinear term (which favors the pinning
of the spin density wave towards the easy-axis), the Hamiltonian 
${\cal H}_{int}$ also contains a `Cooper-pairing' term originating from the 
cosine of the dual field, which
guarantees the presence of quantum fluctuations at
zero temperature. Now, we introduce two Majorana fields
\begin{equation}
\xi_{1\nu}=\frac{\psi_{s\nu}+\psi_{s\nu}^{\dag}}{\sqrt{2}},\hskip 0.5cm 
\xi_{2\nu}=\frac{\psi_{s\nu}-\psi_{s\nu}^{\dag}}{\sqrt{2}i},\hskip 0.5cm 
(\nu=R,L)
\end{equation}
and by using (\ref{un}) we find\cite{Shelton,Egger}
\begin{eqnarray}
\frac{1}{\pi a}\cos\sqrt{4\pi}\Phi_s&=&i(\xi_{1R}\xi_{1L}+
\xi_{2R}\xi_{2L}),\\ \nonumber
\frac{1}{\pi a}\cos\sqrt{4\pi}\Theta_s&=&-i(\xi_{1R}\xi_{1L}-\xi_{2R}\xi_{2L}).
\end{eqnarray}
Refermionization of the spin sector then yields
\begin{eqnarray}
{\cal H}(s)&=&\frac{-i v_F}{2}\sum_{j=1}^2\int dx\ 
(\xi_{jR}\partial_x\xi_{jR}-\xi_{jL}\partial_x\xi_{jL})\\ \nonumber
&+&i\Delta_c\int dx\ \xi_{2R}\xi_{2L}
\end{eqnarray}
Of course, for $m_2=-\Delta_c$\footnote{Including the nonuniversal
number $\gamma<1$ (since ${\vec{n}}^2\sim 2{\gamma}^2=1$) in the Eq.(23), 
the mass $m_2$ will be slightely rescaled as:
$m_2^*=m_2\gamma=-\Delta_c\gamma$.}, the model flows to strong 
couplings rendering
the Majorana field $\xi_2$ massive. The Majorana fermion
$\xi_1$ remains {\it massless}. 

The Hamiltonian ${\cal H}(s)$ shows 
explicitly that
the bosonic mode $\Phi_s$ (or $\Phi^{(1/2)}$ in the non-Abelian language)
decouples into two modes of real (Majorana) fermions (or half-spinons) having 
different spectra. We have a splitting of the
spinon field and it should lead to $\langle\hbox{Tr}\vec{\sigma}\Phi^{(1/2)}
\rangle\not=\hbox{constant}$. 
The spin-singlet real fermions $\xi_1$ remain
free. Let us stress that it contributes to a {\it new} chiral fixed point. 
The specific heat is still linear at low
temperatures, $T\ll \Delta_c$ and comes from gapless spin excitations. Using
the general formula $C_V=\pi CT/3v$\cite{aff2}, we find 
$C_V=\pi T/6v_F$. A free
Hamiltonian of Majorana fermions is ruled by a central charge $C=1/2$
\cite{Tsvelik2}. 

\subsection{Remnant of electronic spin fluctuations}

To compute spin correlation functions, it is
accurate to exploit the well-known correspondence between the 2D Ising model
and 1D Majorana fermions. Both are described by $C=1/2$. We
can define two decoupled Ising models as follows\cite{Itzykson},
\begin{eqnarray}
\label{quatre}
\cos\sqrt{\pi}\Phi_s&=&\mu_1\mu_2,\qquad \sin\sqrt{\pi}\Phi_s=\sigma_1\sigma_2
\\ \nonumber
\cos\sqrt{\pi}\Theta_s&=&\sigma_1\mu_2,\qquad \sin\sqrt{\pi}\Theta_s=\mu_1
\sigma_2
\end{eqnarray}
Bosonic exponents are expressed in terms of the order ($\sigma$) and disorder 
($\mu$) parameters of two Ising models. On the other hand, a theory of a 
massive Majorana fermion field describes long-distance properties of the 
two-dimensional Ising model, the fermionic mass being proportional
to $m\sim (T-T_c)/T_c$. We conclude that ${\cal H}(s)$ is equivalent to two 
decoupled
2D Ising models. The Ising model $(\sigma_1,\mu_1)$ related to $\xi_1$ will be
critical ($T=T_c$), while the Ising model $(\sigma_2,\mu_2)$ related to
$\xi_2$ will be {\it under} criticality (since $m_2=-\Delta_c<0$ we have
$T<T_c$). To perform the complete correspondence, using Eq. (\ref{quatre})
one obtains: $\xi_1\sim\cos\sqrt{\pi}(\Phi_s+\Theta_s)\sim \sigma_1\mu_1$ and
$\xi_2\sim\sin\sqrt{\pi}(\Phi_s+\Theta_s)\sim \sigma_2\mu_2$. 

Since the second Ising model is under criticality, the average of the 
disorder operator
is zero, $\langle\mu_2\rangle=0$. The order operator $\langle\sigma_2\rangle$
has then a finite value, and correlation functions of $\cos\sqrt{\pi}\Phi_s$
and $\cos\sqrt{\pi}\Theta_s$ decay exponentially. From the exact solution of
the 2D Ising model one immediately obtains\cite{Itzykson}, 
\begin{equation}
\langle
\sin\sqrt{\pi}\Phi_s(x)\sin\sqrt{\pi}\Phi_s(x')\rangle\sim\left|x-x'\right|
^{-1/4}
\end{equation}
with the same result for the $\sin\sqrt{\pi}\Theta_s$ operator. The above
proves that the fields $\Phi_s$ and $\Theta_s$ are not pinned: values of
$\sin\sqrt{\pi}\Phi_s$ and $\sin\sqrt{\pi}\Theta_s$ are not fixed to 1. By
virtue of the Heisenberg uncertainty relation, we have simply shown that it
is impossible to completely pin a self-dual field. Note that for
$\lambda_3=0$, scaling dimensions of all these bosonic operators is 1/4. For
$\lambda_3\not =0$, it rescales $1/4\rightarrow 1/8$. More generally, the
operators $\sin\sqrt{2\pi}\Phi_s$ and $\sin\sqrt{2\pi}\Theta_s$ acquire
a `halved' scaling dimension: $\eta=\beta^2/8\pi$ with $\beta=\sqrt{2\pi}$
(in respect to the ground
state $\cos\sqrt{2\pi}\Theta_s$ ($\cos\sqrt{2\pi}\Phi_s$) tends to zero). 

The uniform part of the correlation functions decays exponentially. The nature 
of this new chiral fixed point mainly manifests itself in {\it staggered} 
{\it fusion rules} between spins of conduction electrons:
\begin{equation}
\hskip -0.15cm \hbox{Tr}\sigma^a\Phi^{(1/2)\dag}({\bf
x})\hbox{Tr}\sigma^{b}\Phi^{(1/2)}({\bf
0})\sim\frac{\delta_{ab}}{z^{1/2}}+\epsilon^{abc}(\frac{z}{\bar{z}})^{1/2}J_c^c
\hskip -0.15cm
\end{equation}
with ${\bf x}=(x,\tau)$, ${\bf 0}=(0,0)$ and $z,\bar{z}=x\pm iv_F\tau$. The 
occurrence of the factor $(z/\bar{z})^{1/2}$ is usual and confirms that 
braiding properties of spinons are those of
semions\cite{Bouwknegt}. On the other hand, we emphasize that the halved 
scaled exponent 1/2 in the first term characterizes a new universality class 
for the
S=1/2 Heisenberg chain which is finally equivalent to a C=1/2 critical
theory. In this new chiral fixed point half of the spinon field still
fluctuates. Regarding simply a spinon (spin 1/2 topological object) 
as a domain wall
$(...\downarrow\uparrow\downarrow\downarrow\uparrow\downarrow...)$, the fact
that the electronic 
$\xi_2$ modes acquire a mass (in our language) means physically
that the other half participates in bound states with local moments. 
Taking into account local moment degrees of freedom, 
observable massive excitations are triplets only. Then, we have:
\begin{equation}
\Phi^{(1)}+\xi_1=\Phi^{(1/2)\dag}\Theta(x,\tau)
\end{equation}
$\Phi^{(1)}$ being defined in Eq.(16). This is the main result of 
the present work\cite{exp}. 
\vskip 0.1cm
It
can lead to interesting experimental predictions. A related quantity of 
direct experimental relevance is the NMR relaxation rate
${\cal T}_1$. For large $J_K$, the RKKY regime fails $(C(x)\rightarrow 0)$ 
and the
main contribution comes from the staggered spin-spin correlation functions in
the electron gas. The corresponding susceptibility has the temperature
dependence 
\begin{equation}
\chi(2k_F)\sim T^{2\eta -2}\ \hbox{with}\ 2\eta=1/2
\end{equation}
 Then, we predict
an NMR relaxation rate, $1/{\cal T}_1\propto T^{-1/2}$. This $T$-dependence 
is quite unusual because the underlying magnetic order between electrons
changes the effective field seen by the nuclei. But a related behavior has 
also been 
reported for the so-called dimerized, frustrated and ladder models when the
magnetic field reaches the critical field $H_{c}$ (typically the spin
gap)\cite{Chitra}. The resulting ${\cal T}_1$ is different from that in the 
single impurity case. There,
Friedel oscillations in the unitary limit lead to an NMR rate which in 
contrast increases with the distance\cite{Eggert,Egger1}. 

\section{Fixed point and conclusion}

Now, let us turn to the stability of the chiral phase
when one moves away from $(aL)^2=0$ in the far IR limit, where strong
fluctuations are permitted. There, 
$1-(a\vec{L})^2\sim{\gamma}^2
\rightarrow 0$: the $\lambda_3$ Kondo exchange becomes 
$2k_F$ oscillating, leading to free interchain S=1 
spins in the spectrum $(m_2^*\rightarrow 0)$. However, the
charge sector should not be 
affected because the localization length is sufficiently large, ${\cal
L}_{loc}\gg a$. Integrating out the remnant of massive $\xi_2$ modes,
the leading far IR behavior of the $q=0$ electronic spin component coincides
again with the one in the effective S=1/2 spin chain. 
At small enough
$\lambda_2$, the forward Kondo exchange can be viewed as a weak perturbation
to the chiral fixed point. Starting from order ${\lambda_2}^2$ on, spin flips
contribute:
\begin{equation}
\frac{d\lambda_2}{d\ln L}=\frac{1}{2\pi v_F}{\lambda_2}^2
\end{equation}
The system flows to a purely Kondo state at an energy scale: 
\begin{equation}
m\sim J_K S\exp-2\pi v_F/\lambda_2\ll\Delta_c
\end{equation} 
typically the single-site Kondo temperature. 
Both $\xi_1$ and $\xi_2$ acquire a mass. We do not predict a particular 
lattice enhancement effect\cite{Tsunetsugu}. A similar conclusion has
been obtained in ref.\cite{Fujimoto} (but there is an important mistake in
the recursion law for the Kondo coupling $\lambda_3$). All the spin
correlation functions now decay exponentially. 
Such a
Kondo liquid is another theoretical example of spin-charge 
separation (occurring in the 1D Kondo lattice model) in
agreement with experiments on three dimensional 
Kondo insulators\cite{Aeppli} and previous
numerical results\cite{Shibata}. This can be shown by the
absence of a field-induced metal-insulator transition\cite{Carruzzo}. 
The system remains insulating even when a finite magnetization is induced by
the external field because
$m\ll\Delta_c$. The most typical case is the large-U limit of the 1D
Hubbard model. The charge gap is of order U while magnetic excitations are
gapless. During the magnetization process (occurring for a small field $\sim
t^2/U$), it does not change. 

In summary, there exists an intermediate but still low-energy region where the
chiral fixed point presented here is dominant. It brings new physics in
Kondo ladder systems. Indeed, a spin liquid with both massive
triplet and critical Ising modes takes place. It should be noted that similar
spin spectra have been occurred in other coupled spin chain systems, like
the so-called Kagom\'e lattice
antifferomagnet stripped to its basis\cite{Azaria} or the
two-leg spin ladder with a dimer four-spin interaction\cite{Ners}.
On the other hand,
strong fluctuations among localized spins in the far IR produce
gapless triplet excitations so that the chiral phase becomes unstable. Driven
by the dynamics of localized spins, the system flows to a typical Kondo
fluid. Based on previous numerical results,  the gap value is similar to the
single-impurity Kondo temperature.

The author thanks B. Dou\c cot, M. Fabrizio and T. Jolicoeur for fruitful 
remarks and constructive criticisms, and the theoretical group of ETH for 
stimulating discussions.


\begin{references}
\bibitem{Aeppli} G. Aeppli and Z. Fisk, Comments Condens. Matter Phys.
{\bf 16}, 155 (1992).
\bibitem{Tsunetsugu} H. Tsunetsugu {\it et al.}, Phys. Rev. B {\bf 46}, 3175 
(1992).
\bibitem{Yu} C.C. Yu and S.R. White, Phys. Rev. Lett. {\bf 71}, 3866 (1993).
\bibitem{Tsvelik} A.M. Tsvelik,  Phys. Rev. Lett. {\bf 72}, 1048 (1994).
\bibitem{Zachar} O. Zachar, S.A. Kivelson and V.J. Emery, Phys. Rev. Lett. {\bf
77}, 1342 (1996).
\bibitem{KLH} K. Le Hur, Phys. Rev. B {\bf 58}, 10261 (1998).
\bibitem{Shibata} 
For a recent review see, N. Shibata and K. Ueda, cond-mat/9808301.
\bibitem{Haldane0} F.D.M. Haldane, Phys. Lett. {\bf 93A}, 464 (1983).
\bibitem{Azaria} P. Azaria, C. Hooley, P. Lecheminant, C. Lhuillier, and
A.M. Tsvelik, Phys. Rev. Lett. {\bf 81}, 1694 (1998).
\bibitem{Betouras} J.J. Betouras and S. Fujimoto, Phys. Rev. B {\bf 59}, 529 
(1999).
\bibitem{Rice} E. Daggoto and T.M. Rice, Science {\bf 271}, 618 (1996).
\bibitem{White} S. White and I. Affleck,  Phys. Rev. B {\bf 54}, 9862 (1996).
\bibitem{KLH2} K. Le Hur, Phys. Rev. B {\bf 56}, 14056  (1997).
\bibitem{Haldane} F.D.M. Haldane, J. Phys. C {\bf 14}, 2585 (1981); 
H.J. Schulz, Int. J. Mod. B {\bf 5},
57 (1991); J. Voit, Rep. Prog. Phys. {\bf 58}, 977 (1995).
\bibitem{affleck} I. Affleck, Nucl. Phys. B {\bf 265}, 409 (1986).
\bibitem{Tsvelik2} A.M. Tsvelik, {\it Quantum Field Theory in Condensed Matter
Physics} (Cambridge University Press, 1995).
\bibitem{Bernard} F.D.M. Haldane, Z.N.C. Ha, J.C. Talstra, D. Bernard
and V. Pasquier, Phys. Rev. Lett. {\bf 69}, 2021 (1992).
\bibitem{HS} F.D.M. Haldane, Phys. Rev. Lett. {\bf 60}, 635 (1988);
 B.S. Shastry, Phys. Rev. Lett. {\bf 60}, 639 (1988).
\bibitem{Bouwknegt}  P. Bouwknegt, A. Ludwig and K. Schoutens, 
Phys. Lett. B {\bf 338}, 448 (1994).
\bibitem{BA} Let us define the holon as the excitation associated with
adding or removing a charge $\pm e$ in the system.
\bibitem{Fradkin} E. Fradkin, {\it Field Theories of Condensed Matter
systems} (Addison-Wesley, Reading, MA, 1991), Chap. 5.5.
\bibitem{remark} We have not included the Umklapp operator because it is 
`only' marginal. The Kondo term $\lambda_3$ is more
relevant and produces the localization of the electron
gas at an higher energy scale.
\bibitem{Fujimoto} S. Fujimoto and N. Kawakami, J. Phys. Soc. of Japan
{\bf 66}, 2157 (1997).
\bibitem{Schulz} H.J. Schulz, Phys. Rev. B {\bf 53}, R2959 (1996).
\bibitem{Shelton} D.G. Shelton, A.A. Nersesyan and A.M. Tsvelik,  
Phys. Rev. B {\bf 53}, 8521 (1996).
\bibitem{Egger} R. Egger and A.O. Gogolin, Eur. Phys. J. B {\bf 3}, 281 (1998).
\bibitem{aff2} I. Affleck,  Phys. Rev. Lett. {\bf 56}, 764, 2763 (1986).
\bibitem{Itzykson} C. Itzykson and J.M. Drouffe, 
{\it Statistical Field Theory}, Vol. 1, Chap. 2 (Cambridge University Press, 
1989).
\bibitem{exp} When ${\vec n}$ is assumed to be a constant vector, the same
study implies that both $\xi_1$ and $\xi_2$ become massive. 
The S=1/2 {\it classical} magnetization of the electron gas has 
been completely pinned
by the perfect {\it uniaxial} staggering potential created by the spin array
and physical excitations are massive triplets only 
[Eq.(16)].
\bibitem{Chitra} R. Chitra and T. Giamarchi, Phys. Rev. B 
{\bf 55}, 5816 (1997).
\bibitem{Eggert} S. Eggert and I. Affleck, Phys. Rev. Lett. {\bf 75}, 934
(1995).
\bibitem{Egger1} R. Egger and H. Schoeller,  Phys. Rev. B 
{\bf 54}, 16337 (1996).
\bibitem{Carruzzo} H.M. Carruzzo and C.C. Yu, Phys. Rev. B 
{\bf 51}, 10301 (1995).
\bibitem{Ners} A.A. Nersesyan and  A.M. Tsvelik,  Phys. Rev. Lett. {\bf 20},
3939 (1997).
\end{references}
\end{document}